\documentclass{aa}
\usepackage{graphicx}
\usepackage[varg]{txfonts}
\usepackage{natbib}
 \bibpunct{(}{)}{;}{a}{}{,}    %% natbib format like A&A and ApJ
\begin{document}
   \title{A non-LTE spectral analysis of the $^{3}$He and $^{4}$He
   isotopes\\ in the HgMn star $\kappa$~Cancri\thanks{
   Based on data products from observations made with ESO Telescopes
   at the La Silla Paranal Observatory
   under programme ID 076.B-0055(A).}}

%   \subtitle{}

   \author{Natalia L.~Maza
          \inst{1,}\thanks{Visiting scientist at the Institute for
	  Astro- and Particle Physics, University of Innsbruck.}
          \and Mar\'ia-Fernanda Nieva
          \inst{2,3}
          \and Norbert Przybilla
          \inst{3}
          }

   \institute{Instituto de Ciencias Astron\'omicas, de la Tierra y del
   Espacio (ICATE), Av. Espa\~{n}a 1512 sur, 5400 San Juan,~Argentina
       \and
              Dr. Karl Remeis-Observatory \& ECAP, University of Erlangen-Nuremberg,
Sternwartstr.~7, 96049 Bamberg, Germany
          \and    Institute for Astro- and Particle Physics,
	  University of Innsbruck, Technikerstr. 25/8, 6020 Innsbruck, Austria\\
\email{nmaza@icate-conicet.gob.ar}
%; Maria-Fernanda.Nieva@uibk.ac.at; Norbert.Przybilla@uibk.ac.at}
           %  \thanks{}
             }

   \date{}

% \abstract{}{...}{...}{...}{} 
% 5 {} token are mandatory
 
  \abstract
 {}
%aims heading (mandatory)
 {We present a pilot study on NLTE line-formation computations for the isotopes
 $^3$He and $^4$He in the mercury-manganese star $\kappa$~Cancri.
 The impact of NLTE effects on the determination of isotopic abundances and the
 vertical stratification of helium in the atmosphere is investigated.}
  %methods heading (mandatory)
  {Modern NLTE line-formation computations were employed to
  analyse a high-resolution and high-S/N ESO-VLT/UVES spectrum of $\kappa$\,Cnc.
  The atmospheric parameters were determined from fitting the hydrogen
  Balmer lines and the spectral energy distribution. 
  Multiple \ion{He}{i} lines were investigated, including \ion{He}{i}
  $\lambda$4921\,{\AA} and $\lambda$6678\,{\AA}, which show the widest isotopic
  splits.}
 {Half of the observed \ion{He}{i} lines in the spectrum of
 $\kappa$\,Cnc show significant NLTE strengthening, the effects are strongest in
 the red lines \ion{He}{i} $\lambda$5875\,{\AA} and \ion{He}{i} $\lambda$6678\,{\AA}. 
 NLTE abundances from individual \ion{He}{i} lines are up to a factor of 
 $\sim$3 lower than LTE values. Helium is found to be stratified in
 the atmosphere of $\kappa$\,Cnc. While the LTE analysis indicates a step-like 
 profile of the helium abundance, a gradual decrease with height is indicated by
 the NLTE analysis. A $^3$He/$^4$He ratio of $\sim$0.25-0.30 is found.
 With the available data it cannot be decided whether the two isotopes
 follow the same stratification profile, or not.}
 {This work implies that NLTE effects may be ubiquitous in the
 atmospheres of HgMn stars and may have a significant impact on 
 abundance determinations and the interpretation of the vertical 
 abundance stratification of elements.}


\titlerunning{NLTE spectral analysis of $^{3}$He and $^{4}$He in
$\kappa$\,Cnc}
\authorrunning{Maza et al.}
   \maketitle

%%%%%%%%%%%%%%%%%%%%%%%%%%%%%%%%%%%%%%%%%%%%%%%%%%%%%%%%%%%%%%%%%%%%%%%%%%%%%

\section{Introduction}
Mercury-manganese stars are chemically peculiar (CP) stars of the upper main
sequence, with the line spectra of \ion{Hg}{ii} and \ion{Mn}{ii} 
indicating large overabundances of the two elements. They constitute Preston's 
subgroup CP3 \citep{preston74}. Their effective temperatures range between 
10\,500 and 16\,000\,K, corresponding to spectral types A0 to B6
\citep{Smith1996}. Some other distinctive characteristics of these stars are 
low rotational velocities \citep[$v\sin i\,\le$\,29\,km\,s$^{-1}$,][]{abt1972} 
and overabundances of other elements such as P, Ga, Y, and Pt, with abundance
patterns changing from one star to the other. They can show an
inhomogeneous distribution of some chemical elements over their surface
\citep[first discussed by][]{hubrig1995}, which causes the observed line-profile 
variability in the spectra \citep[e.g.][]{hubrig2011}.
HgMn stars are members of binaries in more than 50\% of the cases, with
periods ranging between 3 to 20 days \citep{gerbaldi1985}.

The chemical peculiarities in the atmospheres of HgMn stars arise
because of their extremely stable atmospheres, 
exposing the atoms and ions to the competitive actions of
gravitational settling and radiative levitation. The observed
abundance peculiarities can in principle be explained within the
framework of atomic diffusion \citep{michaud1979}. Diffusion theory
predicts that the photospheres of all HgMn stars should be deficient in
helium, which was confirmed for instance in a dedicated study by \citet{dworetsky2004}. 
The reason is that helium has only few and weak lines at wavelengths
where the stellar flux is high (longward of the Lyman edge), therefore
gravitational settling dominates.
Dworetsky also found evidence for a vertical helium abundance 
stratification in two HgMn stars, which would also be a consequence of
atomic diffusion in the atmosphere. 

Moreover, helium is a particular case for the occurrence of 
differential effects of diffusion on the {\em isotopes} of an
element. The mass difference between its two isotopes, $^3$He and
$^4$He, is largest among the light elements, with gravitational
settling favouring the heavier isotope.
Some CP stars can therefore show enhanced $^3$He well above the
protosolar $^3$He/$^4$He ratio of (1.66\,$\pm$\,0.05)$\times$10$^{-4}$
\citep{mahaffy1998}. This is most pronounced for the (He-weak) $^{3}$He stars,
which occupy a narrow strip in the $\log T_\mathrm{eff}$--$\log g$
plane between the He-strong B stars and a group of He-weak 
B stars that show no evidence of~$^{3}$He. However, the presence of a
significant amount of $^{3}$He was also suggested for some 
HgMn stars \citep{hartoog1979}. 

\begin{figure*}[t!]
\includegraphics[width=.94\linewidth]{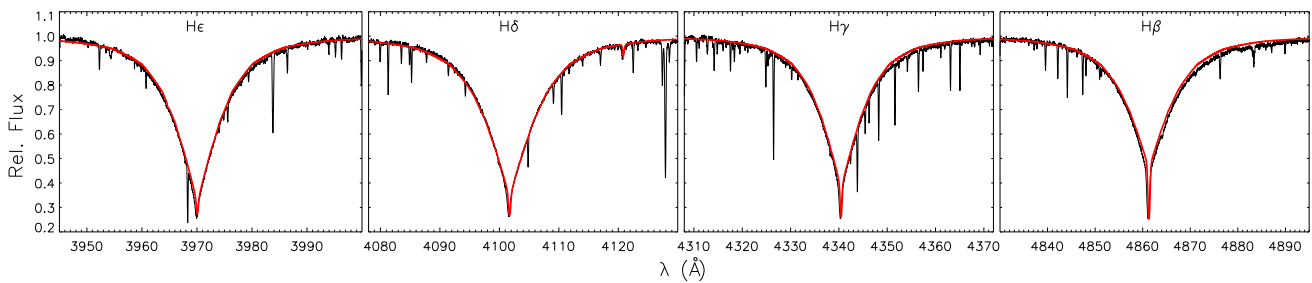}
\caption{Comparison of our global best-fit NLTE synthetic spectrum (red
line) with observed spectral lines of H$\beta$ to H$\epsilon$ (black line)
in $\kappa$\,Cnc.}
\label{fig:hydrogen}
\end{figure*}

Analyses of helium abundances in CP stars (and also for other
elements) have so far mostly been
performed under the assumption of local thermodynamic equilibrium (LTE).
On one hand, this was possibly motivated by the early non-LTE (NLTE) study by
\citet{auer1973}, who found that the blue-violet \ion{He}{i} lines are
described well under the assumption of LTE in the temperature range
below 15\,000\,K. On the other hand, this has also practical reasons
because LTE line-formation can be much easier
implemented than NLTE modelling. However, the quality of the
observations has improved tremendously over the past decades, and we
consider the topic worth to merit a re-visit in an era where quantitative
spectroscopy can be performed at high precision and accuracy. A more
consistent
NLTE modelling of CP stars may therefore help in developing a quantitatively
refined view of diffusion theory, which is required to explain the
variety of observed abundance peculiarities even within
one class of~CP~stars.

We chose the prototype HgMn star \object{$\kappa$\,Cancri}
(HD\,78316, HR\,3623) for a pilot study that aims at re-investigating the
topic based on our latest NLTE models.
This well-known sharp-lined star was reported to be a possible 
$^{3}$He star by \citet{hartoog1979}, who derived an upper limit on the 
$^{3}$He/$^{4}$He ratio of 0.35, more than a thousand times higher than the
protosolar value. The presence of $^{3}$He in the photosphere of 
this star was confirmed by \citet{Dobrichev89} and
\citet{zakharova1996}, concluding on a value of the $^{3}$He/$^{4}$He ratio of 0.35.
These previous works were all based 
on LTE modelling and spectra at lower
resolution than available today. Here, we investigate NLTE effects
and their impact on the determination of (isotopic) abundances and the 
vertical stratification of helium, considering various \ion{He}{i} lines.

%%%%%%%%%%%%%%%%%%%%%%%%%%%%%%%%%%%%%%%%%%%%%%%%%%%%%%%%%%%%%%%%%%%%%%%%%%%%%

\section{Observational data}
The spectrum employed in the present work was observed with {\sc Uves}
\citep[UV-Visual Echelle Spectrograph,][]{Dekkeretal00} on the ESO
VLT/UT2 at Cerro Paranal inChile under the program 076.B-0055(A). 
We extracted the pipeline-reduced spectrum from the ESO Science Archive Facility. 
The spectrum was obtained in dichroic mode, covering
the spectral range $\lambda\lambda$ 3700-10\,200\,{\AA} (with a small
gap in the region of the near-IR calcium triplet),
at resolving power $R$\,=\,$\lambda/\Delta\lambda$\,$\approx$\,110\,000. 
The peak S/N ratio is $\sim$\,270. To normalize of the
spectrum we fitted a spline function to a number of continuum windows.

In addition, (spectro-)photometric data were
adopted from the literature to construct the spectral energy
distribution (SED) of $\kappa$\,Cnc. $UBV$ photometry was adopted from 
\citet{Mermillod91} and $JHK$ data from the Two Micron
All Sky Survey \citep[2MASS,][]{Skrutskieetal06}. Two low-dispersion
spectra that were observed with the
International Ultraviolet Explorer (IUE) using a large aperture 
were extracted from the MAST archive\footnote{\tt
http://archive.stsci.edu/}. The exposures SWP06911 and LWR05873
cover the range from 1150 to 1980\,{\AA} and from 1850 to 3290\,{\AA}. 

\begin{table}
\centering
\caption{Atmospheric parameters}\label{tab:data} 
\vskip -0.1in
\begin{tabular}{lr}
\hline
\hline  
Parameter                &$\kappa$\,Cnc \\
\hline
Sp. Type                 & B8\,III   \\[.5mm]
$V$\,(mag)	         & 5.233    \\
$B-V$\,(mag)             & $-$0.113\\
$E(B-V)$\,(mag)          & 0.038\\[.5mm]
$T_\mathrm{eff}$\,(K)    & 12800$\pm$200  \\
$\log g$\,(cgs)          &  3.70$\pm$0.10   \\   
$\xi$\,(km\,s$^{-1}$)    & 0+1      \\
$v\sin i$\,(km\,s$^{-1}$)& 6$\pm$2        \\
$\zeta$\,(km\,s$^{-1}$)  & 4$\pm$2       \\
\hline
\end{tabular}
\end{table}

%%%%%%%%%%%%%%%%%%%%%%%%%%%%%%%%%%%%%%%%%%%%%%%%%%%%%%%%%%%%%%%%%%%%%%%%%%%%%

\section{Model calculations}
We employed a hybrid NLTE approach for our line-formation
calculations, which has been successfully used before for quantitative
analyses of main-sequence OB stars
\citep{nieva2007,nieva2008}, their evolved progeny
\citep{przybilla2006a}, and subdwarf B stars
\citep{Przybillaetal06b}. 
Model atmospheres were computed with the code {\sc Atlas9}
\citep{kurucz1993b}, which assumes plane-parallel
geometry\footnote{The ratio of atmospheric thickness (geometrical
height from the outer rim to the optical continuum-forming layers from
the {\sc Atlas9} model) over stellar radius (determined from evolutionary 
mass and the surface gravity) is $\sim$1\%. This means that sphericity effects
are negligible here, despite the star's classification as giant.},
chemical homogeneity, and hydrostatic, radiative, and local
thermodynamic equilibrium (LTE). A value of triple solar metallicity
was adopted for the model atmosphere calculations for $\kappa$\,Cnc,
which is found to reproduce the low-resolution UV spectrum overall well.

Then, NLTE line-formation computations were performed with the
codes {\sc Detail} and {\sc Surface} (Giddings 1981;
Butler \& Giddings 1985). {\sc Detail} calculates atomic-level
populations by solving the coupled radiative transfer and statistical equilibrium
equations, and {\sc Surface} computes the formal solution using realistic 
line-broadening functions. The following model atoms were employed:
H \citep{przybilla2004}, \ion{He}{i}
\citep{przybilla2005}, and, for some supplementary calculations, \ion{O}{i/ii}
\citep[updated]{przybilla2000,becker1988}. To facilitate NLTE
calculations for the $^3$He isotope in addition to the usual $^4$He, a \ion{$^3$He}{i} 
model atom in analogy to that of $^4$He by \citet{przybilla2005} was realized, taking
into account isotopic line shifts as measured by \citet{fred1951}. Both isotopes 
were treated simultaneously to account for the overlap of lines and continua.  

Our approach allows LTE spectrum synthesis to be performed by
running {\sc Surface} only on top of the {\sc Atlas9} model. In this
case, LTE level populations are computed while all other model
ingredients remain unchanged, which facilitates assessing the impact of NLTE effects.

%%%%%%%%%%%%%%%%%%%%%%%%%%%%%%%%%%%%%%%%%%%%%%%%%%%%%%%%%%%%%%%%%%%%%%%%%%%%%

\section{Atmospheric parameter determination}
$\kappa$\,Cnc is a known single-lined spectroscopic binary
(SB1) star. The system has been resolved among others by
\citet{Schoelleretal10}, who found a
$\sim$2.5\,mag difference in the $K$ band between the primary and the
(later-type) secondary. Because the brightness ratio 
increases towards shorter wavelengths, we do not expect the
available spectrum to be contaminated by any significant second light.

\begin{figure}[t!]
\centering
\includegraphics[width=.99\linewidth,height=5.5cm]{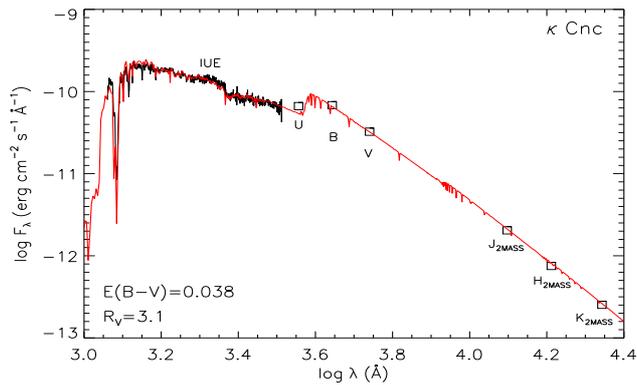}
\caption{Comparison of the {\sc Atlas9} model flux (red line) 
for our adopted atmospheric parameters with the observed SED (black
line and boxes). Values of colour excess and the ratio of
total-to-selective extinction employed in the de-reddening of the
observations are indicated.}
\label{fig:SED}
\end{figure}

\onlfig{
\begin{figure}[t!]
\includegraphics[width=.99\linewidth]{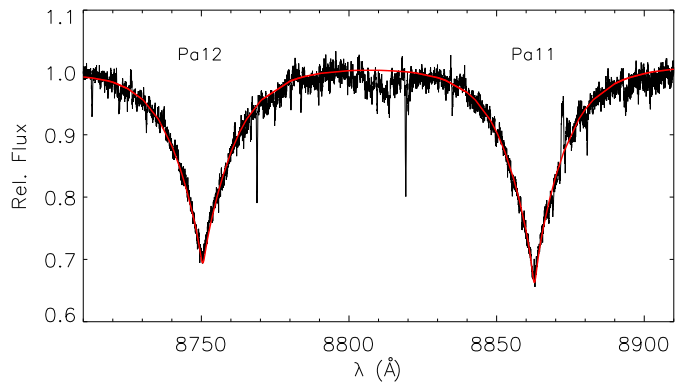}
\caption{Comparison of our global best-fit NLTE synthetic spectrum
(red line) with the observed Paschen lines Pa11 and Pa12 (black line)
in $\kappa$\,Cnc. 
}
\label{fig:paschen}
\end{figure}
}
\onlfig{
\begin{figure*}[bh!]
\includegraphics[width=.99\linewidth]{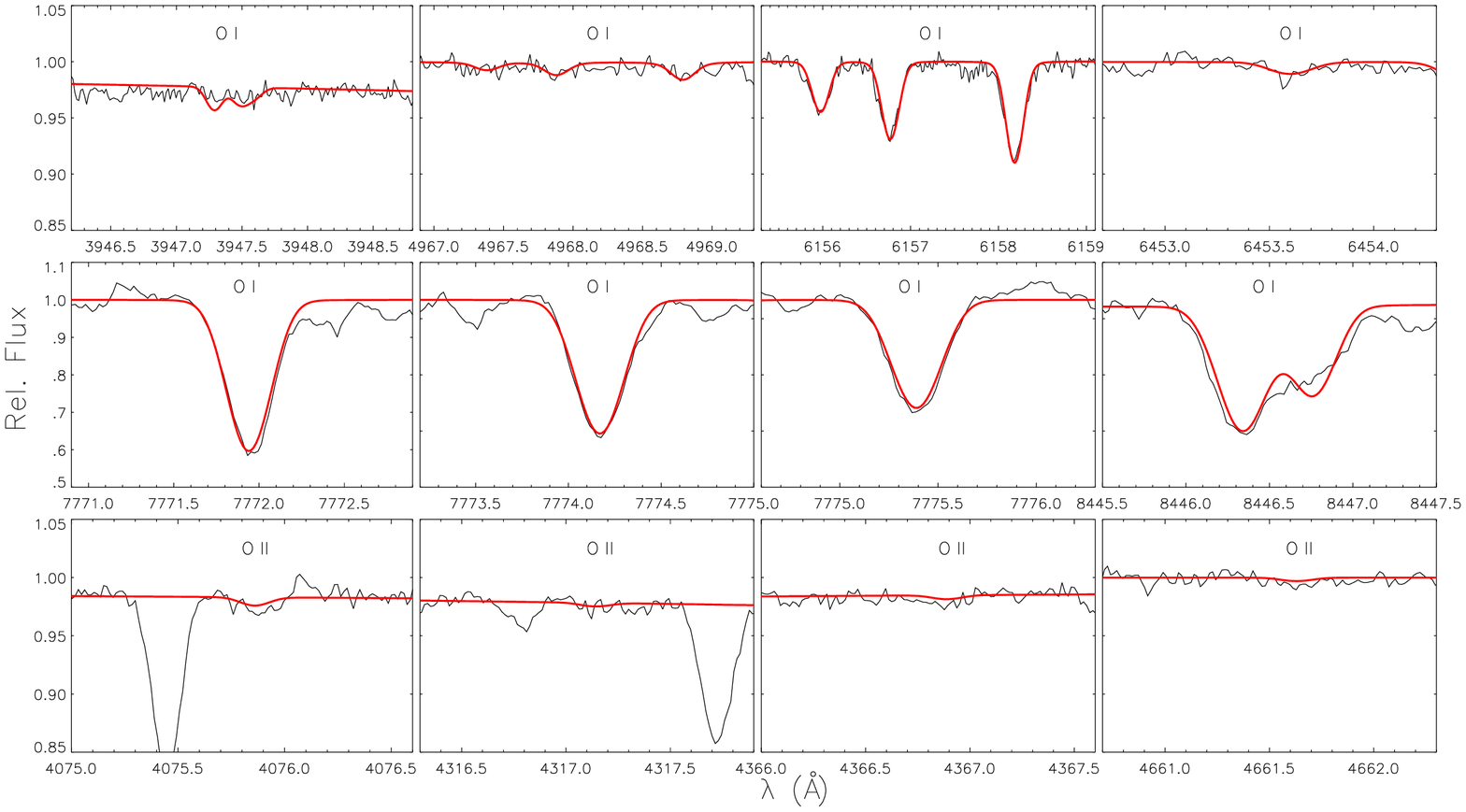}
\caption{Comparison of our global best-fit NLTE synthetic spectrum
(for $\log$\,(O/H)\,+\,12\,$=$\,8.40, red) with the observed
\ion{O}{i/ii} lines (black lines) in $\kappa$\,Cnc. 
}
\label{fig:oxygen}
\end{figure*}
}

Therefore, the analysis of the observed spectrum followed a similar philosophy as previously 
employed by us for studies of normal B-type stars in \citet{nieva2007,nieva2012}.
We used the hydrogen Balmer lines and the SED as principal
indicators to determine the effective temperature
$T_\mathrm{eff}$ and the surface gravity $\log g$. 
Supplementary indicators were two hydrogen Paschen lines and
the ionization balance of \ion{O}{i/ii}. A fine-tuning of the
atmospheric parameters was achieved by multiple iterations aiming at
reproducing all indicators simultaneously. At this stage of the
analysis, an approximate average value for the helium abundance in
the line-formation region was determined, which had to be assumed
to be homogeneous throughout the entire atmosphere because of the basic
assumptions made in {\sc Atlas9}. Note that this introduces no restrictions to
the further discussion, as helium behaves as a trace element in our case.
A zero microturbulent
velocity $\xi$ was adopted, which facilitates reproducing the
weak and strong oxygen lines for one abundance value simultaneously. The (radial-tangential)
macroturbulent velocity $\zeta$ and projected rotational velocity $v\sin i$ 
were also determined based on fits to the oxygen lines. The
finally adopted atmospheric parameters together with a few
supplementary data are summarised in Table~\ref{tab:data}.
Our $T_\mathrm{eff}$ is significantly lower (by up to $\sim$1000\,K in
the extremes) and the $\log g$
slightly lower than in earlier LTE analyses
\citep[e.g.][]{RoLa90,DwBu00,Adelmanetal04,bailey2013,Takedaetal14}. The
higher temperatures are not compatible with the observed SED.
On the other hand, our results agree excellently well with the atmospheric
parameters adopted by \citet{zakharova1996} and \citet{Mazaetal11}.

Figures~\ref{fig:hydrogen} and \ref{fig:SED} show a comparison of a 
global (by-eye) best-fit NLTE spectrum with H$\beta$ to H$\epsilon$ and the fit
of the {\sc Atlas9} model flux to the observed SED.
For the latter, the photometric
data were converted into fluxes using zeropoints of \citet{Besselletal98}
for Johnson photometry and of \citet{Heberetal02} for
the 2MASS photometry. All observed fluxes were de-reddened
using an interstellar reddening law according to \citet{Cardellietal89},
adopting a colour excess $E(B-V)$ as indicated in Table~\ref{tab:data}
and a ratio of total-to-selective extinction
$R_V$\,=\,$A_V/E(B-V)$\,=\,3.1.
Overall, good agreement between model and observation is achieved.
This also includes the confirmation of the atmospheric parameter
determination by the secondary indicators, the Paschen lines P$_{11}$ and P$_{12}$ 
(the only useful lines from the near-IR region, see the online
Fig.~\ref{fig:paschen}) and the \ion{O}{i/ii} NLTE
ionization balance (see the online Fig.~\ref{fig:oxygen}). 
The latter is fulfilled for a homogeneous oxygen abundance of
$\log$\,(O/H)\,+\,12\,$=$\,8.40, without any indication
for a depth-dependent abundance stratification of this element.

\begin{figure}[t!]
\centering
\includegraphics[width=.99\linewidth,height=6.8cm]{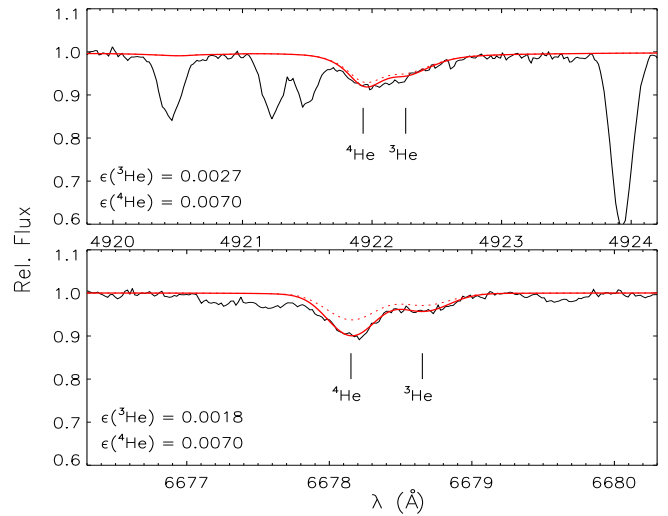}
\caption{Comparison of our NLTE spectrum synthesis (red
lines) with observation for the two lines with the widest
isotopic separation (black lines).
Synthetic LTE spectra computed for the same isotopic helium
number fractions are also shown (dotted red line).
The positions of the line centres of the isotopic components are
indicated.}
\label{fig:3he}
\end{figure}

\begin{figure}[t!]
\centering
\includegraphics[width=.99\linewidth,height=6.8cm]{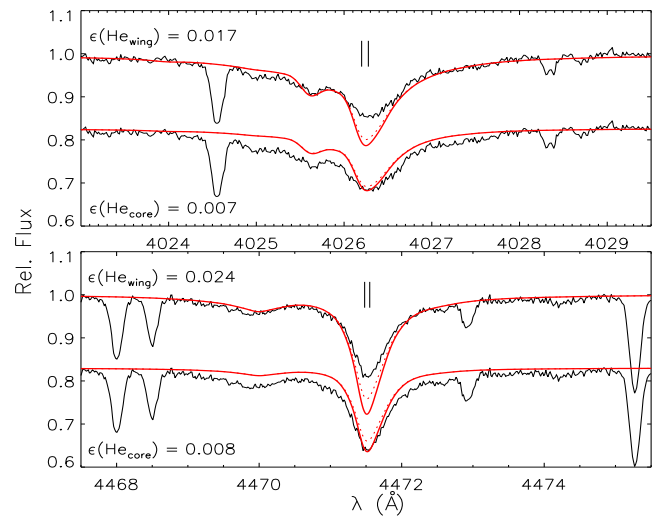}
\caption{Comparison of our two-component NLTE fits
with the strongest observed diffuse \ion{He}{i} lines. 
The upper curves in each panel visualise the fit to the line
wings, the lower ones the fit to the line core. See Fig.~\ref{fig:3he}
for the description of the line encoding. The two vertical lines indicate 
the degree of the isotopic shift in the two cases.}
\label{fig:hewings}
\end{figure}

%%%%%%%%%%%%%%%%%%%%%%%%%%%%%%%%%%%%%%%%%%%%%%%%%%%%%%%%%%%%%%%%%%%%%%%%%%%%%

\begin{figure}
\includegraphics[width=.99\linewidth]{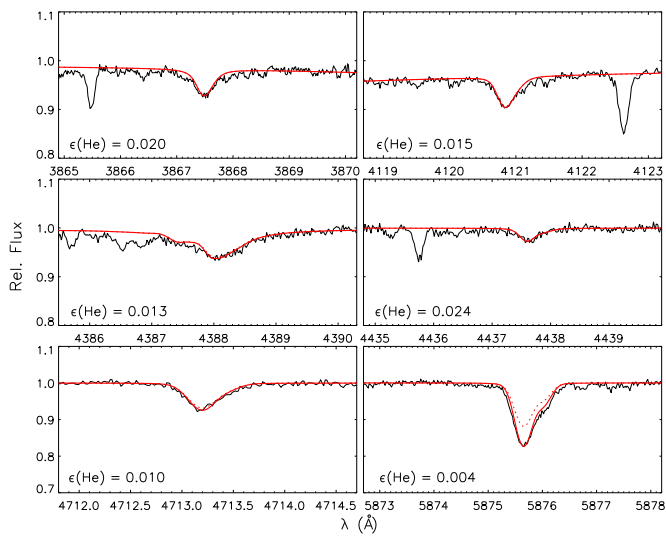}
\caption{Comparison of our NLTE spectrum synthesis
with all other \ion{He}{i} lines that were analysed here. Shown are individual best
fits. Lines are encoded in the same way as in Fig.~\ref{fig:3he}.}
\label{fig:herest}
\end{figure}

\section{Helium abundance analysis}
The visual inspection of the helium spectrum of $\kappa$\,Cnc shows
two peculiarities. First, an asymmetry of \ion{He}{i}
$\lambda$4921.9\,{\AA} and a resolved second component redwards of 
\ion{He}{i} $\lambda$6678.1\,{\AA} indicate the presence of the $^3$He
isotope in the stellar atmosphere (see Fig.~\ref{fig:3he}), confirming
the detection of \citet{hartoog1979} at much higher spectral resolution
\citep[cf.~also][]{zakharova1996}.
Second, the strongest among the diffuse \ion{He}{i} lines, at 
$\lambda$4026.2\,{\AA} and $\lambda$4471.5\,{\AA}, show shallow cores
in combination with unusually broad wings (see
Fig.~\ref{fig:hewings}). This is a clear sign
that the element is not homogeneously distributed throughout the
atmosphere, but instead shows a vertical abundance stratification,
which is unusual in HgMn stars and has been reported for only a
few objects so far \citep{dworetsky2004,CaHu07}. 

The quantitative analysis of the \ion{He}{i} spectrum uses line-profile 
fits to the observed (largely) unblended features (\ion{He}{i}
$\lambda\lambda$ 4009.3\,{\AA}, 5015.7\,{\AA}, and 5047.7\,{\AA} are
strongly blended with metal lines). The individual best fits from the
NLTE modelling to the two lines with resolved isotopic contributions are shown in
Fig.~\ref{fig:3he}. For comparison, LTE profiles for the same derived
helium abundances are also shown. While NLTE strengthening is weak for
the \ion{He}{i} $\lambda$4922\,{\AA} lines (note a weak blend by
\ion{Fe}{ii} $\lambda$4922.19\,{\AA}, which is not accounted for in
the modelling), \ion{He}{i} $\lambda$6678.1\,{\AA}
shows pronounced NLTE effects. A helium abundance higher
by a factor of almost 3 is indicated by the quantitative LTE analysis in the latter case.
The $^3$He and $^4$He number fractions
$\varepsilon$(He) derived from the NLTE and LTE analysis are summarised in
Table~\ref{isotopic-abundances}, together with the Rosseland optical
depths $\tau_\mathrm{ross}$ at which the line cores are formed. The abundance ratio
$^3$He/$^4$He is about 0.25 to 0.30 at the atmospheric depths covered
by the observations, slightly lower than the value of 0.35 derived by 
\citet{zakharova1996}.

\begin{figure}
\centering
\includegraphics[width=.98\linewidth]{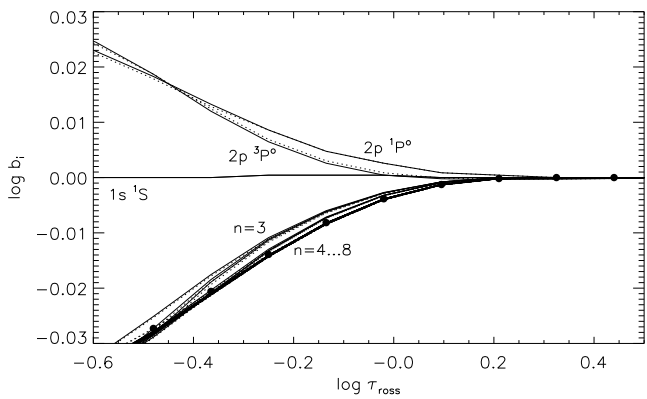}
\caption{Departure coefficients $b_i$ of the $^4$He (full lines)
and the $^3$He levels (dotted lines) as a function of optical depth
for a model with $\varepsilon$($^4$He)\,=\,0.008 and
$\varepsilon$($^3$He)\,=\,0.002. Level designations or principal quantum
numbers $n$ of the levels are indicated. The dots mark the run of the \ion{He}{ii}
ground-state departure coefficient.}
\label{fig:departures}
\end{figure}

The isotopic shifts are much smaller in all other helium lines, such
that only a combined abundance from both isotopes can be derived
(we kept the $^3$He/$^4$He ratio fixed to 0.3, consistent with
the value range determined above during
our modelling). For the strongest diffuse \ion{He}{i} lines
we performed a two-component fit (Fig.~\ref{fig:hewings}). One model was
tuned to reproduce the line core, the other to reproduce the line wings.
The derived abundances are summarised in Table~\ref{he-abundances} in
analogy to Table~\ref{isotopic-abundances}. Data for the wings of
the lines were evaluated 2\,{\AA} redwards of the line core.
In the cores of the two lines, NLTE abundances are smaller than
derived in LTE, by about 15\% and 40\%.
The wings are unaffected by NLTE effects.
Individual best fits to all other \ion{He}{i} lines that were analysed
here are shown
in Fig.~\ref{fig:herest}, with the results summarised in
Table~\ref{he-abundances}. Typically, NLTE effects are weak in these
mostly weak lines. Strong effects are present only in \ion{He}{i}
$\lambda$5875.6\,{\AA}, implying that the NLTE abundances are lower
than the LTE values by a factor~of~almost~3.
\paragraph{NLTE effects.}
Departure coefficients $b_i$\,=$n_i/n_i^*$ (Zwaan definition, the 
$n_i$ and $n_i^*$ are the NLTE and LTE level population numbers) 
for the helium energy levels are displayed in Fig.~\ref{fig:departures} 
as a function of optical depth. \ion{He}{i} is the main ion at the
temperatures found in the visible layers of the atmosphere of
$\kappa$\,Cnc. The population of the \ion{He}{i} ground state shows
practically no deviations from detailed equilibrium. The levels with principal
quantum number $n$\,=\,2 are overpopulated relative to LTE values 
(the 2$s$\,$^{1,3}$S levels are not shown in Fig.~\ref{fig:departures}
for clarity, as they show qualitatively and quantitatively a very
similar behaviour as the 2$p$\,$^{1,3}$P$^\circ$ levels). 
This is because radiative (the 2$s$\,$^{3}$S state is metastable) 
and collisionally induced decays to the ground state are inefficient, 
while the level populations are constantly fed by 
the recombination cascade from higher levels, which drains the
populations of these. In consequence, all higher
levels are underpopulated compared with the LTE values. They closely follow the
behaviour of the \ion{He}{ii} ground state. Detailed balance is
approached for all levels at continuum formation depths (at $\log
\tau_\mathrm{ross}$\,$\approx$\,0.2-0.3) and below, the NLTE
population numbers differ at most by a few percent from LTE values in
the line-formation region (see Tables~\ref{isotopic-abundances} and
\ref{he-abundances}). The differences in the
behaviour of the departure coefficients between $^3$He and $^4$He levels
are negligible.
\begin{table}
\centering
\caption{Isotopic helium abundances}\label{isotopic-abundances} 
\vskip -0.1in
\begin{tabular}{lrrr}
\hline
\hline  
$\lambda$\,({\AA})	&$\log\tau_\mathrm{ross}$ & $\varepsilon$(He)$_\mathrm{NLTE}$  &$\varepsilon$(He)$_\mathrm{LTE}$\\
\hline
\ion{$^4$He}{\sc i} 4921.9 & $-$0.061 & 0.0070 & 0.0087\\
\ion{$^3$He}{\sc i} 4922.3 &    0.001 & 0.0027 & 0.0032\\
\ion{$^4$He}{\sc i} 6678.1 & $-$0.362 & 0.0070 & 0.0185\\
\ion{$^3$He}{\sc i} 6678.7 & $-$0.256 & 0.0018 & 0.0032\\
\hline
\end{tabular}
\end{table}
\begin{table}
\centering
\caption{Line-by-line abundances of \ion{He}{i}.}\label{he-abundances} 
\vskip -0.1in
\begin{tabular}{lrrr}
\hline
\hline  
$\lambda$\,({\AA})	&$\log\tau_\mathrm{ross}$ & $\varepsilon$(He)$_\mathrm{NLTE}$  &$\varepsilon$(He)$_\mathrm{LTE}$\\
\hline
3867.5      &    0.179 & 0.020 & 0.020\\
4026.2 core &    0.045 & 0.007 & 0.008\\
4026.2 wing &    0.213 & 0.017 & 0.017\\
4120.8      &    0.105 & 0.015 & 0.016\\
4387.9	    &    0.098 & 0.013 & 0.015\\
4437.6      &    0.142 & 0.024 & 0.024\\
4471.5 core & $-$0.146 & 0.008 & 0.011\\
4471.5 wing &    0.167 & 0.024 & 0.024\\
4713.2      & $-$0.001 & 0.010 & 0.012\\
5875.6      & $-$0.373 & 0.004 & 0.011\\
\hline
\end{tabular}
\end{table}
All the \ion{He}{i} lines that were analysed here have either the 2$p$\,$^1$P$^\circ$ 
or the 2$p$\,$^3$P$^\circ$ level as the lower transition level and 
upper levels with a higher principal quantum number. Therefore, NLTE
strengthening of the lines can occur as the consequence of overpopulated lower and
underpopulated upper levels, depending on the formation
depth. The (stronger) red \ion{He}{i} lines are more sensitive to NLTE
effects also because of the pronounced response of the line-source function
to variations of the departure coefficients in the
Rayleigh-Jeans limit \citep[see e.g.][]{przybilla2004}. Even
small differences in level populations on the percent-level 
can therefore result in fundamentally altered line strengths.

%%%%%%%%%%%%%%%%%%%%%%%%%%%%%%%%%%%%%%%%%%%%%%%%%%%%%%%%%%%%%%%%%%%%%%%%%%%%%

\section{Discussion}
\begin{figure}
\includegraphics[width=.99\linewidth,height=5.8cm]{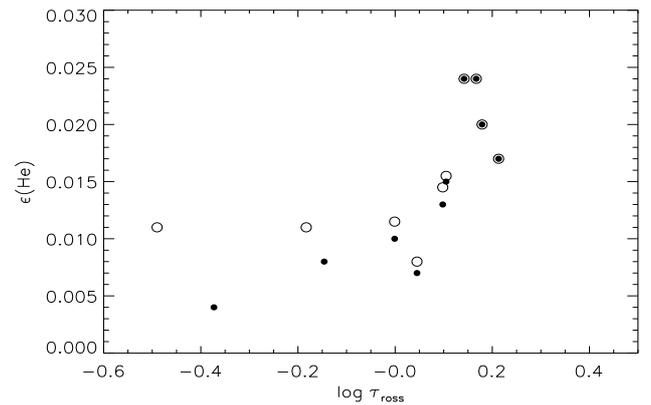}
\caption{Helium number fraction %$\varepsilon$(He) 
as a function of Rosseland optical depth %$\tau_\mathrm{ross}$ 
in the atmosphere of $\kappa$\,Cnc (NLTE results: dots, LTE: circles).}
\label{fig:stratification}
\end{figure}

Using abundances from lines of different strength -- which sample
the line-formation region to a different extent --, one can reconstruct the abundance profile,
that is, the vertical abundance stratification of an element in the atmosphere.
The run of the helium number fraction $\varepsilon$(He) with 
optical depth in the atmosphere of $\kappa$\,Cnc
is shown in Fig.~\ref{fig:stratification}. The LTE analysis indicates
a step-like helium stratification, dropping in the region
$\log \tau_\mathrm{ross}$\,$\approx$\,0.2 to 0.1
from a high abundance of $\varepsilon$(He)\,$\approx$\,0.25 
to a constant value of about half of that farther outward in the
atmosphere. On the other hand, the stratification derived from the NLTE
analysis implies a gradual decrease of the helium abundance with
decreasing optical depth, which is a qualitatively different behaviour.
The NLTE abundances are lower by almost a
factor 3 at the lowest optical depths that can be traced. Note that
while the formation regions of the weaker lines agree well
when computed under LTE and NLTE conditions, they can  
differ by up to $\sim$0.1\,dex in $\log \tau_\mathrm{ross}$ for the stronger lines.
The LTE computations indicate a formation of the line cores farther out in the
atmosphere.

It would be highly interesting to know whether the two helium isotopes
follow the same stratification profile, or if differential
effects prevail. Unfortunately, with data from only two lines
this cannot be decided here with confidence.

Finally, it may be appropriate here to comment on NLTE effects in CP
stars in the general context, based on the present findings. The
helium lines in $\kappa$\,Cnc are weak because of the rather low
$T_\mathrm{eff}$ for populating the excited levels and because of the
low helium abundance that is due to gravitational settling. Nevertheless, 
NLTE effects are found for all but
the weakest \ion{He}{i} lines, leading to changes in the derived
abundances relative to an LTE analysis by up to a factor $\sim$3.
On the other hand, most of the metals show large overabundances
because of radiative levitation, yielding spectral lines much stronger
than usually found in normal stars. The formation regions for
the strong lines extend to much lower optical depths than investigated
here, possibly giving rise to much stronger NLTE effects than
encountered in the present work. More NLTE studies for CP stars are
encouraged because this may have a significant effect on the results of
elemental abundance determinations and on the deduced
abundance stratifications. 

\begin{acknowledgements}
We thank V.~Schaffernroth for helping us to improve the normalization
of the observed spectrum. We also thank our anonymous referee for several
suggestions that helped to improve the manuscript.  
NLM acknowledges a CONICET postdoc stipend and additional
financial support by CONICET under the `Programa de financiamiento
parcial de estad\'ias breves en el exterior para becarios
postdoctorales', and MFN an equal opportunity FFL stipend from the 
University of Erlangen-Nuremberg. Financial support by the International
Relations Office to a research stay of NLM at University of Innsbruck is
gratefully acknowledged.
\end{acknowledgements}

\end{document}